\documentclass{aa}  
\usepackage{xcolor}
\usepackage{txfonts,comment, amsmath, longtable}
\usepackage[colorlinks,linkcolor={blue},citecolor={blue}]{hyperref}
\usepackage{lineno}
\usepackage{ulem}
\definecolor{blue}{rgb}{0.,0.,0.5}
\definecolor{gray}{rgb}{0.,0.5,0.}
\definecolor{darkblue}{rgb}{0,0,0.5}
\definecolor{orange}{rgb}{1.,0.5,0.}
\definecolor{darkgreen}{rgb}{0,0.5,0}

\definecolor{dylan}{rgb}{0.78, 0.08, 0.52}

\begin{document} 

   \title{Raising the observed metallicity floor with a 3D non-LTE analysis of SDSS$\,$J102915.14+172927.9\thanks{Based on observations obtained at ESO Paranal Observatory, programme 286.D-5045,  P.I. P. Bonifacio.}}
  
  \authorrunning{C. Lagae et al.}
  \titlerunning{Raising the observed metallicity floor}
  
   \author{C.\,Lagae\inst{\ref{SU}},
          A.\,M.\,Amarsi\inst{\ref{UU}},
          L. F. Rodríguez Díaz\inst{\ref{AU}},
          K.\,Lind\inst{\ref{SU},\ref{A3D}},
          T.\,Nordlander\inst{\ref{A3D},\ref{ANU}},
          T.\,T.\,Hansen\inst{\ref{SU}}
          \and
          A.\,Heger\inst{\ref{A3D},\ref{MU}}
          }

   \institute{\label{SU}Department of Astronomy, Stockholm University,
              Albanova University Center, 106 91 Stockholm, Sweden\\
              \email{cis.lagae@astro.su.se}
              \and 
              \label{UU}Theoretical Astrophysics,
              Department of Physics and Astronomy, Uppsala University,
              Box 516, SE-751 20 Uppsala, Sweden
              \and 
              \label{AU}Stellar Astrophysics Centre, Department of Physics and Astronomy, Aarhus University, Ny Munkegade 120, DK-8000 Aarhus C, Denmark
              \and 
              \label{A3D}ARC Centre of Excellence for All Sky Astrophysics in 3 Dimensions (ASTRO 3D), Australia
              \and 
              \label{ANU}Research School of Astronomy and Astrophysics, Australian National University, Canberra, ACT 2611, Australia
              \and
              \label{MU}School of Physics and Astronomy, Monash University, Clayton, Vic 3800, Australia
             }

   \date{Accepted 2 March 2023; Received 23 December 2022}

 
  \abstract
   {The first stars mark the end of the cosmic dark ages, produce the first heavy elements, and set the stage for the formation of the first galaxies. Accurate chemical abundances of ultra metal-poor stars ($[\mathrm{Fe}/\mathrm{H}]<-4$) can be used to infer properties of the first stars, and thus the formation mechanism for low-mass second generation stars in the early universe. Spectroscopic studies have shown that most second generation stars are carbon-enhanced. A notable exception is SDSS J102915.14+172927.9, the most metal-poor star known to date, largely by virtue of the low upper limits of the carbon abundance reported in earlier studies.}   
   {We reanalyse the composition of SDSS J102915.14+172927.9 with the aim to provide improved observational constraints on the lowest metallicity possible for low-mass star formation and constrain the properties of its Population III progenitor star.}
   {We developed a tailored three dimensional (3D) model atmosphere for SDSS J102915.14+172927.9 with the \texttt{Stagger}-code, making use of an improved surface gravity estimate based on the Gaia DR3 parallax. Snapshots from the model were used as input in the radiative transfer code \texttt{Balder} to compute 3D non-local thermodynamic equilibrium (non-LTE) synthetic spectra. These spectra were then used to infer abundances for Mg, Si, Ca, Fe and Ni, and upper limits on Li, Na and Al. Synthetic 3D LTE spectra were computed with \texttt{Scate} to infer the abundance of Ti and upper limits on C and N.}
   {In contrast to earlier works based on 1D non-LTE corrections to 3D LTE results, we are able to achieve ionisation balance for \ion{Ca}{I} and \ion{Ca}{II} when employing our consistent 3D non-LTE treatment. The elemental abundances are systematically higher than those found in earlier works. In particular, [Fe/H] is increased by $0.57\,\mathrm{dex}$, and the upper limits of C and N are larger by $0.90\,\mathrm{dex}$ and $1.82\,\mathrm{dex}$, respectively.}  
   {We find that Population III progenitors with masses $10-20\,\mathrm{M}_\sun$ exploding with energy $E\lessapprox3\cdot10^{51}\,\mathrm{erg}$ can reproduce our 3D non-LTE abundance pattern. Our 3D non-LTE abundances are able to better constrain the progenitor mass and explosion energy as compared to our 1D LTE abundances. Contrary to previous work, we obtain higher upper limits on the carbon abundance that are ``marginally consistent'' with star formation through atomic line cooling, and as such, prevent strong conclusions about the formation mechanism of this low mass star.
   }


   \maketitle
%

\section{Introduction}\label{Sec:introduction}
The first stars (Population III stars, hereafter Pop III stars) in our universe formed directly from Big Bang nucleosynthesis products: H, He and traces of Li. These stars were the source of the first enrichment of the interstellar and intergalactic medium, and set the stage for the formation of the first galaxies and started the epoch of reionisation \citep{Bromm11}. 

Despite their importance, the details of star formation and the resulting initial mass function (IMF) of Pop III stars are still uncertain. Initial studies suggested that the lack of metals in star-forming clouds caused the absence of an efficient cooling mechanism and suppressed the formation of low-mass stars in the early universe. Hydrodynamic simulations based on cooling by hydrogen molecules predicted stellar masses larger than $>100\,\mathrm{M}_\sun$ \citep{Bromm11}. More recent studies that incorporate longer timescales, higher resolution, and improved physics such as subsonic turbulence \citep{Clark11b}, fragmentation of accretion disks \citep{Greif11,Clark11a}, protostellar radiation feedback \citep{Hosokawa11,Hosokawa16,Hirano14}, multiplicity \citep{Stacy13}, and a combination of these \citep{Susa13,Susa14,Stacy16} allow stars to form with initial masses of several solar masses from $\lesssim1\,\mathrm{M}_\odot$ up to $1000\,\mathrm{M}_\sun$. The lack of direct observations of Pop III stars and expected scarcity of long-lived low-mass ($M\lesssim 0.8\,\mathrm{M}_\odot$) Pop III stars in the galaxy \citep{Hartwig15,Magg19}, however, makes it difficult to directly constrain these predictions.

One way to indirectly study the properties of the first stars is to look at metal-poor second-generation stars (Population II stars). These stars are born from the gas that the Pop III stars enrich; moreover, studies claim that stars with metallicity of up to $[\mathrm{Fe}/\mathrm{H}]<-3$ could, in principle, be formed from a cloud enriched by a single supernova \citep{Tominaga07b,Nomoto13,Keller14,Frebel15}. Using a semi-analytical model of early universe star formation, \cite{Hartwig18} found that $40\%$ of stars with $-6\lessapprox[\mathrm{Fe}/\mathrm{H}]\lessapprox-4$ are enriched by only one nucleosynthesis event (mono-enriched). Therefore, by comparing the chemical compositions of ultra metal-poor stars (UMP, $\mathrm{[Fe/H]}<-4$, \citealt{Beers05}) to theoretical yields of first-star core-collapse supernovae, it is, in principle, possible to constrain the explosion properties and IMF of the first stars -- at least for single stars, although some questions about asymmetric supernovae and dilution remain \citep{Ezzeddine19,Magg20}. These types of analyses indicate that the progenitors of Pop II stars typically have stellar masses in the range of $10-100\,\mathrm{M}_\sun$ \citep{Lai08,Heger10,Joggerst10,Ishigaki14,Ishigaki18,Tominaga07b,Tominaga14,Placco15}. 

A related problem concerns the transition from Pop III star formation to low-mass star formation that follows the currently known IMF peaking at $\sim0.2-0.3\,\mathrm{M}_\sun$ \citep{Kroupa02,Chabrier03,Bastian10}. Two models have been proposed for this transition to predominantly low-mass star formation: atomic fine-structure line cooling \citep{Bromm03} and dust-induced fragmentation \citep{Schneider03,Omukai05, Ji14}. Atomic fine-structure line cooling can occur if the gas is enriched beyond critical abundances of \ion{C}{II} and \ion{O}{I}, often approximated as $Z_\mathrm{crit}/Z_\sun\sim10^{-3.5}$. On the other hand, dust-induced fragmentation which requires dust formation in the ejecta from first star supernovae, can already operate at critical abundances that are a factor $10-100$ smaller: $Z_\mathrm{crit}/Z_\sun\sim10^{-5}$ \citep{Omukai05,Schneider12a}. 

In the last decade, multiple large spectroscopic surveys have expanded our sample of ultra metal-poor stars allowing us to make a first classification \citep{Beers85,Beers05,Christlieb08}. One result of this work is that the fraction of carbon-enhanced metal-poor (CEMP, $[\mathrm{C}/\mathrm{Fe}]>0.7$; \citealt{Aoki07} and \citealt{Beers05}) stars increases for decreasing metallicity \citep{Placco14b,Arentsen22}.  Below $[\mathrm{Fe}/\mathrm{H}]<-4.5$ we almost exclusively observe CEMP-stars. Due to their carbon-enhanced nature, these stars all obey the critical abundance necessary for atomic line cooling. 

There is one notable exception: SDSS J102915.14+172927.9, hereafter ``SDSS J102915+172927'' \citep{Caffau11}. This is the star with the lowest metal-mass fraction $Z$, currently known, in large parts due to a low upper limit on its carbon abundance: $[\mathrm{C}/\mathrm{H}]\le-4.5$. Its low carbon abundance puts the star below the critical abundance for metal-line cooling, providing support for the dust-induced formation model \citep{Caffau12,Schneider12b,Klessen12}.

Inferences about the Pop III progenitors and about the Pop II formation mechanism, however, rely on stellar chemical abundances that are determined via high-resolution spectroscopy. These analyses are strongly sensitive to the approximations employed when modelling the synthetic spectra. The abundance offsets introduced by the simplifying assumptions of one dimensional (1D) hydrostatic atmospheres and local thermodynamic equilibrium (LTE) have been shown to be particularly severe at lower metallicities \citep{Amarsi16b,Amarsi22,Bergemann12,Bergemann19,Ezzeddine17}.

One source of these discrepancies is that, contrary to 3D simulations wherein convection arises naturally, in 1D hydrostatic models convection must be treated in an approximate way, for example by the mixing-length theory \citep{Bohm1958}. The 1D models also cannot make predictions on other intrinsic 3D hydrodynamic effects such as granulation and cooling by adiabatic expansion. The first 3D models of metal-poor stellar atmospheres showed that the absence of adiabatic cooling in 1D models leads to a severe overestimation of the surface temperature stratification \citep{Stein98,asplund05}. Hence, in LTE, 3D synthetic spectral lines of neutral species that are formed in these surface layers are often stronger than their 1D counterparts. Equally important, non-LTE effects for minority species subject to overionisation are expected to be larger in the steeper temperature gradients of metal-poor stars \citep{Bergemann12}. To mitigate the problems, without attempting fully consistent 3D non-LTE synthesis, estimates of non-LTE and 3D effects have often been made separately as in the case for SDSS J102915+172927 \citep{Caffau11,Caffau12}. The non-LTE effects, however, in fact vary across the surface due to horizontal inhomogeneities \citep{Asplund03, Nordlander17a}. Hence attempting to correct for 3D and non-LTE effects separately, often leads to new abundance offsets. \citet{Nordlander17a} already showed that doing a full 3D non-LTE spectral synthesis of SMSS0313-6708, the most iron-poor star currently known, can significantly change abundance estimates. The full 3D non-LTE analysis increased the upper limit on the Fe abundance for this star to $[\mathrm{Fe}/\mathrm{H}]<-6.53$ as compared to the previous non-LTE upper limit $[\mathrm{Fe}/\mathrm{H}]<-7.52$ computed using averaged 3D ($\langle3\mathrm{D}\rangle$) model atmospheres \citep{Bessel15}. Additionally, recent work on C and O \citep{Amarsi16a}, Li \citep{Wang21}, and Fe \citep{Amarsi16b, Amarsi22} also demonstrated significant 1D LTE $-$ 3D non-LTE abundance corrections for metal-poor stars.

In light of recent advancements in the field of modelling 3D model atmospheres and 3D non-LTE spectral synthesis, we perform for the first time a fully consistent 3D non-LTE abundance analysis of the most metal-poor star SDSS J102915+172927. In $\S$\ref{Sec:method} we present the model atmosphere, radiative transfer and model atoms used to produce 3D non-LTE synthetic spectra. The resulting abundances are presented in $\S$\ref{Sec:Abundances} and discussed in $\S$\ref{Sec:discussion}; our conclusions given in $\S$\ref{Sec:conlusion}.

\section{Method}\label{Sec:method}

\subsection{Observational data and stellar parameters}\label{Sec:StellarParams}
Observational spectra of SDSS J102915+172927 ($G=16.5$, \citealt{Gaia20}) were obtained by the programme 286.D-5045 and principle investigator P.~Bonifacio, using the UVES spectrograph \citep{Dekker00} mounted on the VLT. More information about the instrument settings and other details of the observations can be found in \cite{Caffau12}. For this work, we downloaded the reduced science spectra containing 14 individual exposures of $3005\,\mathrm{s}$ each from the ESO advanced data product archive\footnote{Url: \url{http://archive.eso.org/eso/eso_archive_main.html}}. We find a signal-to-noise ratio per pixel of approximately 35 at 650 nm for each individual exposure, which agrees with the values found by \cite{Caffau12}. Each exposure was aligned manually by adding a radial velocity shift such that the cores of the H Balmer lines and the \ion{Ca}{II}-doublet corresponded to the correct central wavelength. Subsequently, co-adding all exposures resulted in the final spectrum with a signal-to-noise ratio per pixel of $S/N\approx20$ in the blue region around $3500\,\AA$, $\approx70$ around the green region $5200\,\AA$ and $\approx35$ in the red region at $8500\,\AA$ of the spectrum. 

\citet{Caffau12} determined the stellar effective temperature ($T_{\rm\!eff}$) from the $(g-z)$ colour obtained by \cite{Ludwig08}, resulting in a value of $5811\pm150\,\mathrm{K}$. For the surface gravity, it was still unclear at the time of the original analysis whether SDSS J102915+172927 is a dwarf or a subgiant star. \citet{Caffau12} determined a gravity of $\log g=4.0\pm0.5\,\mathrm{dex}$, but not excluding a higher gravity of $\approx 4.5\,\mathrm{dex}$. With the release of an improved parallax by Gaia DR2 $\varpi=0.734\pm0.07\,\mathrm{mas}$, \cite{Bonifacio18} resolved the gravity issue by comparing metal-poor isochrones from A.~Chieffi (Priv.\ Comms.) to their newly computed absolute $V$-magnitude using Gaia DR2 data, concluding that SDSS J102915+172927 is a dwarf star. \cite{Sestito19} computed $T_\mathrm{\!eff} = 5764\pm60\,\mathrm{K}$ and $\log g=4.7\pm0.1\,\mathrm{dex}$ by fitting MESA ischochrones, with $\mathrm{[Fe/H]}=-4$, to observations using Gaia DR2 astrometric and photometric data. The estimated error of using isochrones with a fixed metallicity for stars that are more metal poor than $\mathrm{[Fe/H]}=-4$ is expected to be small and accounted for by adding an additional error of $0.01~\mathrm{mag}$ to the photometric uncertainities \citep{Sestito19}. Using the parallax from Gaia DR3 $\varpi=0.648\pm0.06\,\mathrm{mas}$, Sestito (Priv.\ Comms.) arrived at $T_\mathrm{\!eff} = 5811\pm51\,\mathrm{K}$ and $\log g=4.68\pm0.1\,\mathrm{dex}$. This effective temperature compares well to the original derived value whereas the surface gravity is $0.7\,\mathrm{dex}$ higher but corresponds to SDSS J102915+172927 being a dwarf star. In this work we used the latest values derived by Sestito (Priv.\ Comms.) using Gaia DR3 data. 

\begin{table*}[t]
\begin{center}
    \caption{Summary of the stellar parameters of SDSS J102915+172927 obtained in previous literature studies and for the hydrodynamic models used in this work.}
    \def\arraystretch{1.5}

    \begin{tabular}{ cccc}
     \hline
     \hline
      & $T_\mathrm{\!eff}$ [K] & $\log g$ [dex] & [Fe/H]\\ \hline
      \cite{Caffau12} & $5811 \pm 150$ & $4.0\pm 0.5$ & $-4.89 \pm 0.10$ \\
      \cite{Sestito19} $\&$ Gaia DR2 & $5764 \pm 56$ & $4.69 \pm 0.1$ & $\dots$ \\
      F. Sestito$^2$ $\&$ Gaia DR3 & $5811 \pm 51$ & $4.68 \pm 0.1$ & $\dots$ \\   
      \texttt{Stagger} 3D model atmosphere & $\approx 5810$ & $4.70$ & $-4$ \\
      \texttt{MARCS} 1D model atmosphere & 5811 & 4.68 & $-4$ \\
     \hline

    \end{tabular}
\label{t:stellar parameters}
\end{center}
\footnotesize{$^2$ Private communication}\\
\end{table*}

\subsection{Model atmosphere}
The tailored 3D ``box-in-a-star'' model atmosphere was computed for this work by using the radiative-hydrodynamic \texttt{Stagger}-code \citep{Nordlund95,Nordlund09}, with later improvements by its user community \citep{Magic13b,Collet18}. \texttt{Stagger} solves the fluid conservation equations (mass, energy, momentum) on a staggered Eulerian mesh and radiation is included as heating and cooling terms in the energy conservation equation. The size of the box is chosen such that it contains roughly ten granules and covers the top of the convective zone, superadiabatic region, photosphere and upper layers of the atmosphere. The surface gravity is fixed since the box only represents a small region of the star. The radiative transfer is solved in LTE, where the scattering contribution is included in the total extinction, using an approximate opacity binning method \citep{Nordlund82,Skartlien00,Ludwig13,Collet18} with 12 bins. Since computing the full monochromatic solution in 3D is computationally unfeasible, all wavelength points are sorted in 12 bins according to their formation depth $\tau_\mathrm{Ross}(\tau_\lambda=1)$. For each bin a mean opacity $\kappa_i$, containing both continuum and line opacity, is computed that is subsequently used in the calculation of the mean intensity and radiative heating rate:
\begin{equation}
    q_\mathrm{rad} = 4\pi\rho\sum_i \kappa_i(J_i - B_i)\,\,,
\end{equation}
with density $\rho$, $J_i$ the mean intensity for bin $i$ and $B_i$ the Planck function at the local gas temperature. More details of the implementation in \texttt{Stagger} can be found in \cite{Magic13b}. The Equation of State (EOS) is a customised version from \cite{Mihalas88} constructed by \cite{Trampedach13}, which includes the following species: H, He, C, N, O, Ne, Na, Mg, Al, Si, S, Ar, K, Ca, Cr, Fe, Ni, $\mathrm{H}_2$, and $\mathrm{H}_2^+$. Continuous absorption and scattering coefficients are taken from \cite{Hayek10} and line opacities from \cite{Gustafsson08}.

Surface gravity, metallicity, bottom boundary entropy and a fiducial hyper-viscosity parameter are the only free parameters in the \texttt{Stagger} simulations, with effective temperature being an emergent property. As such, the surface gravity was fixed to the value selected in $\S$\ref{Sec:StellarParams} and metallicity at $\mathrm{[Fe/H]} = -4$ with alpha-enhancement $\mathrm{[\alpha/Fe]}=0.4$ typical for stars at these low-metallicities \citep{Mashonkina17}. The bottom boundary entropy was modified iteratively in order to obtain the desired effective temperature of $\sim5811\,\mathrm{K}$. The H to He abundance ratio is fixed at solar value while the other elemental abundance ratios with respect to H are scaled down from their solar values, for example $[\mathrm{C}/\mathrm{H}]=-4$ and $[\mathrm{N}/\mathrm{H}]=-4$ using solar abundances from \cite{Asplund09}. A summary of the input and emergent stellar parameters of the computed model are given in Table \ref{t:stellar parameters} together with literature estimates. The model itself is set on a $240\times240\times240$ Cartesian grid that is more spatially refined near the surface to resolve the steep photospheric temperature gradient. Information about the numerical details of the code can be found in \cite{Magic13b} and \cite{Collet18}.

\begin{figure}[ht]
  \centering
  \resizebox{3.5in}{!}{\includegraphics{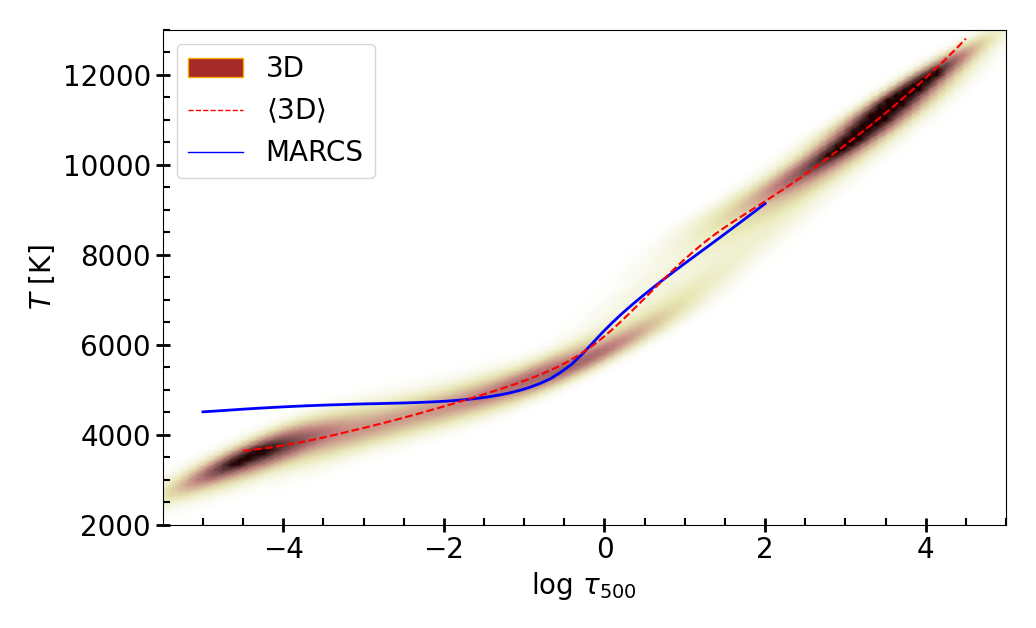}}
\caption[]{Temperature heatmap of the \texttt{Stagger} model with its corresponding average $\langle 3\mathrm{D}\rangle$ of all vertical columns over optical depth (\textsl{red dashed}). The 1D \texttt{MARCS} temperature stratification is shown in \textsl{blue}.  } 
\label{fig:atmos}
\end{figure}

In addition to the tailored 3D model atmosphere, a \texttt{MARCS} atmosphere, interpolated from the \texttt{MARCS} grid \citep{Gustafsson08}, was used to compute 1D abundances. The 1D plane-parallel atmosphere has a metallicity of $[\mathrm{Fe}/\mathrm{H}]=-4$ with similar $\alpha$, C, N, and O abundances as the \texttt{Stagger} model and microturbulence of $1\,\mathrm{km/s}$. The exact value of the microturbulence was relaxed for the post-processing spectrum synthesis described in $\S$\ref{Sec:spectralModelling}. A comparison between the 1D \texttt{MARCS} and 3D \texttt{Stagger} model is made in Fig. \ref{fig:atmos}, where the temperature stratification of both is displayed together with the average stratification of the 3D model ($\langle\mathrm{3D}\rangle$). This average is calculated by taking the mean temperature of all vertical cross-sections at each depth point in optical depth space. It showcases the impact of adiabatic cooling in 3D metal-poor atmospheric models in the upper atmosphere where the temperature structure differs significantly compared to the 1D model. The 3D model has a steeper temperature gradient at the photosphere. Although the $\langle\mathrm{3D}\rangle$-model is an improvement over the 1D model as it accurately describes the temperature stratification of the full 3D model, it loses information on the surface granulation and accompanying impact on spectral line formation.

\subsection{Spectral modelling}\label{Sec:spectralModelling}

\begin{table*}[]
\begin{center}
    \caption{1D (non-)LTE abundance ($A(\mathrm{X})$, see Eq.~\ref{eq:Ax}) sensitivity to the uncertainties on $T_\mathrm{\!eff}$, $\log g$ and $v_\mathrm{mic}$.}
    \def\arraystretch{1.5}
    \begin{tabular}{ rrrrrrr} 
     \hline
     \hline
     Species & $T_\mathrm{\!eff}+100\,\mathrm{K}$ & $T_\mathrm{\!eff}-100\,\mathrm{K}$ & $\log g+0.2\,\mathrm{dex}$ & $\log g-0.2\,\mathrm{dex}$ & $v_\mathrm{mic}+0.5\,\mathrm{km/s}$ & $v_\mathrm{mic}-0.5\,\mathrm{km/s}$ \\
     \hline
     \ion{Mg}{I} & 0.08 & $-0.09$ & $-0.02$ & 0.05 & $-0.01$ & 0.01 \\
     \ion{Ca}{I} & 0.09 & $-0.09$ & $-0.01$ & $0.05$ & $-0.01$ & 0.01 \\
     \ion{Ca}{II} & $-0.04$ & $-0.12$ & $-0.03$ & $-0.12$ & $-0.06$ & 0.08 \\
     \ion{Si}{I} & 0.07 & $-0.10$ & $-0.01$ & 0.01 & $-0.01$ & 0.01 \\
     \ion{Fe}{I} & 0.10 & $-0.11$  & 0.01 & 0.05 & $-0.02$ & 0.03 \\
     \ion{Ti}{II} & 0.07 & $-0.05$ & 0.08 & $-0.06$ & $-0.07$ & 0.09 \\
     \ion{Ni}{I} & 0.11 & $-0.10$ & 0.01 & 0.02 & $-0.04$ & 0.07 \\ \hline
     C & 0.15 & $-0.15$ & $-0.05$ & $0.05$ & $\approx 0$ & $\approx 0$ \\
     N & 0.15 & $-0.15$ & $-0.15$ & $0.05$ & $\approx 0$ & $\approx 0$ \\
     \ion{Li}{I} & 0.08  & $-0.07$ & 0.01 & 0.03 & $\approx 0$ & $\approx 0$ \\
     \ion{Na}{I} & $0.07$ & $-0.06$ & $0.01$ & $0.02$ & $\approx 0$ & $\approx 0$ \\
     \ion{Al}{I} & $0.07$ & $-0.08$ & $-0.01$ & $0.04$ & $\approx 0$ & $\approx 0$\\
     \hline

    \end{tabular}
\label{t:AbundCorrections}
\end{center}
\end{table*}

The radiative transfer code \texttt{Balder} was used to compute 1D and 3D non-LTE synthetic line profiles for the elements for which we have suitable atomic models: Si \citep{Amarsi17}, Li \citep{Wang21}, Ca, Mg \citep{Asplund21}, Na \citep{Lind11}, Al \citep{Nordlander17b} and Fe \citep{Amarsi16b,Amarsi22,Lind17}. The code itself is based upon the original framework of \texttt{Multi3D} \citep{Leenaarts09}. \cite{Amarsi16a,Amarsi16b} expanded the original code, in particular with an improved opacity package \texttt{BLUE}, change in angle quadrature and implementation of frequency parallelisation. 

\texttt{Balder} solves the statistical equilibrium for user-specified trace elements using the multilevel approximate lambda iteration pre-conditioning method of \cite{Rybicki92} on a Cartesian grid. The assumption is made that any deviations from LTE do not impact the temperature stratification or the densities of other elements. The Equation of State, background, and line opacities are treated by the \texttt{BLUE} package \citep{Amarsi16b}, in which the latter is precomputed on a grid of wavelength, density and temperature. To decrease the computational cost of the 3D calculations, the resolution of the \texttt{Stagger} model was reduced to 60 grid points in the horizontal and 101 points in the vertical direction. \cite{Nordlander17a} showed that this mesh reduction impacts the derived abundances by at most $0.03\,\mathrm{dex}$. For the 3D spectral synthesis, we sampled the temporal variation of the atmosphere by computing line strengths from three different snapshots of the \texttt{Stagger} model. These were chosen so as to maximise the difference in $T_\mathrm{\!eff}$ and, therefore, the difference in resulting line profiles. We find mean abundance variations between snapshots of no more than $0.02\,\mathrm{dex}$, which is in line with similar work done by \cite{Nordlander17a}. 

For the other elements that we do not expect large deviations from non-LTE (\ion{Ti}{II}; \citealt{Mallinson22}), or for which we do not have the atomic and molecular data needed to solve the statistical equilibrium equations (\ion{Ni}{I}, CH and NH), we compute 1D and 3D LTE synthetic spectra using \texttt{Scate} \citep{Hayek11}. The necessary atomic line data (oscillator strengths, excitation potential and damping parameters) are taken from \cite{Kurucz16} for \ion{Ti}{II} and \cite{Kurucz08} for \ion{Ni}{I}, using the VALD database \citep{Piskunov95,Ryabchikova15}.

A range of synthetic spectra were calculated with varying abundances ($\Delta=0.2\,\mathrm{dex}$) for the elements: \ion{Li}{I}, CH, NH, \ion{Na}{I}, \ion{Al}{I}, \ion{Mg}{I}, \ion{Si}{I}, \ion{Ca}{I}, \ion{Ca}{II}, \ion{Ti}{II}, \ion{Fe}{I} and \ion{Ni}{I}, also listed in Table \ref{t:abundances}. These were interpolated to the equivalent width ($W_\lambda$) measured in the observed spectrum. The latter is found by fitting one or more, in the case of blended lines, Gaussians to the observed line profiles. An overview of the identified spectral lines and corresponding equivalent widths is given in Table \ref{t:EWobs}. \cite{Caffau12} report the equivalent widths for a selection of lines in their Table 4, all of which agree with our values within their error. For the elements with no visible lines in the UVES spectrum (Li, Na, Al, CH, and NH), upper limits were estimated using either Cayrel's formula \citep[for Li, Na and Al]{Cayrel88,Cayrel04} or a reduced $\chi^2$ statistical test (CH and NH). In the second case we applied artificial broadening to the synthetic spectra to simulate instrumental broadening with a resolution of $38\,000$. Further details on the upper limit determination are given in the respective results section of \ion{Li}{} ($\S$\ref{sec:resultsLi}) and C ($\S$\ref{sec:CHNH}).

\begin{figure}[ht]
  \centering
  \resizebox{3.6in}{!}{\includegraphics{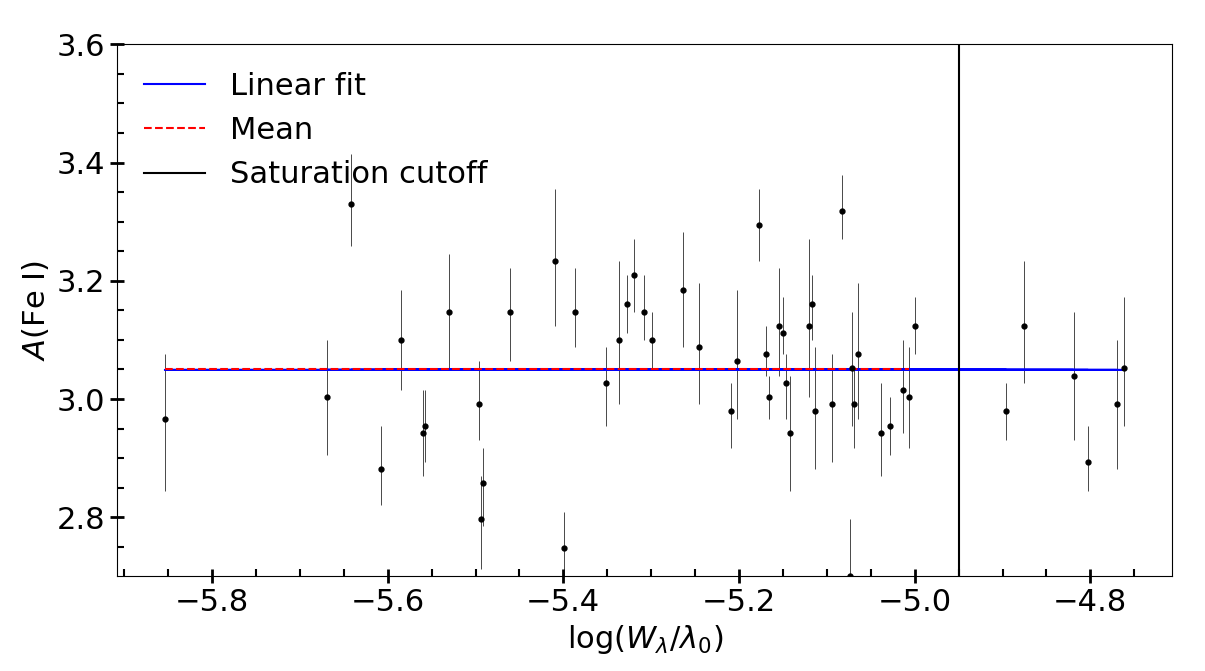}}
\caption[]{1D non-LTE \ion{Fe}{I} abundances for each visible line in the spectrum interpolated to a microturbulence of $v_\mathrm{micro}=1.48\,\mathrm{km/s}$. The \textsl{blue line} shows the best linear fit through the data while the \textsl{red dashed line} denotes the mean non-LTE abundance. The \textsl{black solid line} denotes the cutoff in reduced equivalent width between saturated and non-saturated lines: $\mathrm{log}[W_\lambda/\lambda_0]<-4.95$. } 
\label{fig:Fe_excit}
\end{figure}

Spectrum synthesis calculations in 1D contain microturbulent velocity $v_\mathrm{mic}$ as an extra free broadening parameter to approximate the convective velocity fields. This free parameter is usually determined by flattening the abundances of individual \ion{Fe}{I} and \ion{Fe}{II} lines as a function of reduced equivalent width $W_\lambda/\lambda_0$. Since we only detect \ion{Fe}{I} in the observed spectrum, we computed 1D (non-)LTE \ion{Fe}{I} abundances for three different values of microturbulence $v_\mathrm{mic}=1,\,1.5\,\mathrm{and}\,2\,\mathrm{km/s}$ to obtain a final best fitting value of $v_\mathrm{mic}=1.48\pm0.25\,\mathrm{km/s}$. The result of this analysis is shown in Fig. \ref{fig:Fe_excit} where all detected \ion{Fe}{I} lines are used to determine $v_\mathrm{mic}$. This value is consistent with the empirical analysis performed by \cite{Frebel13}, for a star with $\log g=4.7$. In all subsequent spectral synthesis computations a value of $v_\mathrm{mic}=1.5\,\mathrm{km/s}$ was used.

The stipulated abundance errors originate from manually placing the continuum and from the noise of the spectrum. It is calculated from Cayrel’s formula \citep{Cayrel88,Cayrel04}, which gives an error on the equivalent width that propagates through to the abundance error, with a separate contribution from the continuum placement. In addition, there is an error coming from the uncertainties on $T_\mathrm{\!eff}$, $\log g$ and $v_\mathrm{mic}$, which are computed by shifting one stellar parameter at a time in the \texttt{MARCS} atmosphere and performing an abundance analysis as described above. The error is then equal to the abundance offset between the new model and the model with correct stellar parameters. For example, the \ion{Mg}{I} abundance increases by $0.08~\mathrm{dex}$ when computed using a \texttt{MARCS} atmosphere with a $T_\mathrm{eff}$ that is $100~\mathrm{K}$ hotter. These results are summarised in Table \ref{t:AbundCorrections}. Finally, the total error becomes:
\begin{equation}
    \sigma^2_\mathrm{tot} = \sigma^2_\mathrm{Cayrel} + \sigma^2_\mathrm{continuum} + \sigma^2_{T_\mathrm{\!eff},\log g,v_\mathrm{mic}}\,\, 
\end{equation}
with $\sigma_\mathrm{continuum} = \frac{1}{S/N}\cdot W_\lambda$.

\section{Results}\label{Sec:Abundances}
All abundance results are summarised in Table \ref{t:abundances} in the form:
\begin{equation}\label{eq:Ax}
    A(\mathrm{X}) = \mathrm{log}(x) + A(\mathrm{H})\,\,,
\end{equation}
where $x$ is the number fraction of element X and $A(\mathrm{H})=12$. Whenever an abundance is converted to the solar normalised abundance: $[\mathrm{X}/\mathrm{H}] = A(\mathrm{X}) - A(\mathrm{X})_\sun$, we used the solar abundances from \cite{Asplund21}. 

\subsection{Lithium}\label{sec:resultsLi}
The $\ion{Li}{I}$ line at $6707.6\,\AA$ is not detected in the UVES spectrum such that we could only determine an upper limit. We used Cayrel's formula \citep{Cayrel88}, similar to \cite{Caffau12}, to determine a $3\sigma$ upper limit. The resulting 3D non-LTE upper limit, $A(\mathrm{Li})<1.06\pm0.05$, is slightly higher than the original value, $A(\mathrm{Li})_\mathrm{Caffau+12}<0.9$, but still well below the Spite plateau, $A(\mathrm{Li})_\mathrm{Spite}\sim2.2$ \citep{Spite82,Sbordone10,Melendez10}, implying that the star has undergone significant Li depletion, as expected based on its effective temperature.

\subsection{Carbon and nitrogen}\label{sec:CHNH}
Both C and N are undetected in the UVES spectrum, but the quality of the spectrum allows the calculation of robust upper limits. To obtain these, we computed 1D and 3D LTE synthetic spectra of the CH G-band at $4300\,\AA$ and the NH-band at $3360\,\AA$ for a wide range of abundances $A(\mathrm{C})=3.6-5.6$ and $A(\mathrm{N})=2.4-6.0$ in steps of $\Delta=0.2\,\mathrm{dex}$. The upper limit was determined using a reduced $\chi^2$-statistic where we `fitted' the synthetic spectral bands to the UVES spectrum. The null hypothesis is that our featureless observed spectrum can be fully described by the synthetic spectrum, meaning that both are indistinguishable. The upper limit is then defined as the synthetic spectrum with abundance $A(\mathrm{C})$ that does not statistically describe the observed featureless spectrum.

We start by masking out the blended $\ion{Ti}{ii}$ and $\ion{Fe}{i}$ line in the NH- and CH-band, respectively, and compute the reduced $\chi^2_\nu$ for each synthetic spectrum:
\begin{equation}
    \Bar{\chi}^2_\nu = \frac{1}{\nu}\sum_\lambda \Big(\frac{F_\mathrm{obs}(\lambda) - F_\mathrm{synth}(\lambda)}{\sigma_\mathrm{continuum}(\lambda)}\Big)^2\,\,,
\end{equation}
with $\nu$ the degrees of freedom equal to the number of wavelength points minus 1 and $\sigma_\mathrm{cont}=S/N^{-1}$. For a significance level of $\alpha=0.05$ we can look up the corresponding mean of the $\chi^2$-distribution: $\chi^2_{_\nu,\alpha}/\nu$ \citep{Bognar21}, where the probability that a randomly chosen $\chi^2$ from the distribution function will be greater than $\chi^2_{_\nu,\alpha}$ is equal to the significance level, in this case $5\%$. It follows that the null hypothesis is rejected for synthetic models that have $\Bar{\chi}^2_\nu > \chi^2_{_\nu,\alpha}/\nu$, either because the synthetic spectrum does not properly model the observed spectrum or due to a statistically improbable excursion of probability $5\%$. This means that the upper limit on the CH and NH abundance is equal to the synthetic spectrum for which $\Bar{\chi}^2_\nu = \chi^2_{_\nu,\alpha}/\nu$, as any model with a higher abundance do not statistically describe the featureless spectrum we observe. A visualisation of the $\Bar{\chi}^2_\nu$ values used to determine the upper limits for C and N is shown in Figure \ref{fig:chi2_results}.

\begin{figure}[ht]
  \centering
  \resizebox{3.5in}{!}{\includegraphics{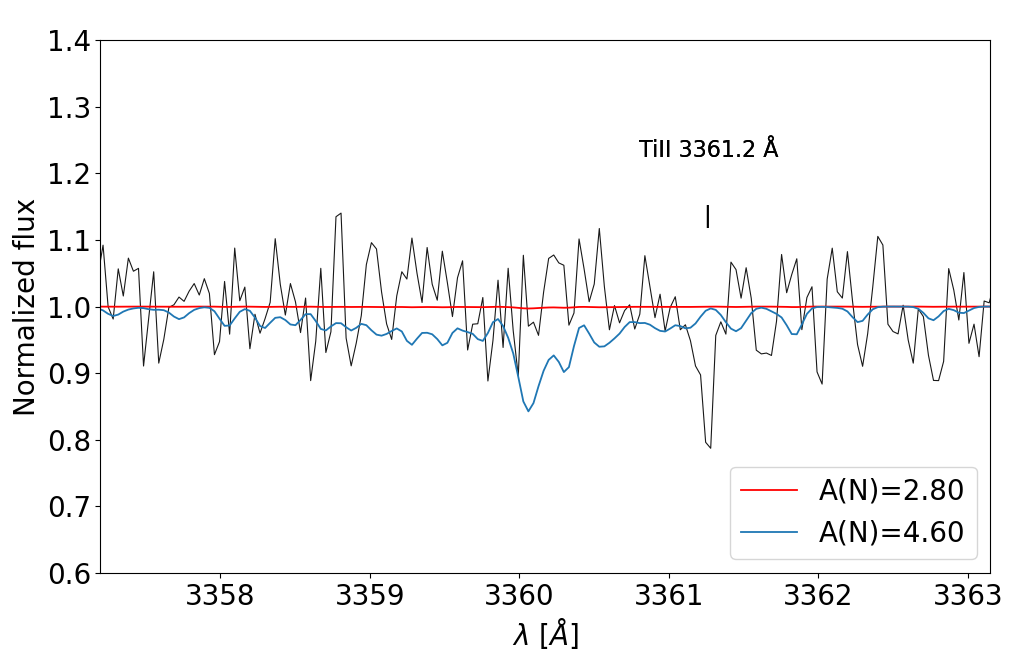}}
\caption[]{The UVES spectrum (black) zoomed in on the NH-band at $3360\,\AA$. The synthetic spectrum corresponding to the upper limit derived in this work is shown in blue, while the upper limit from \cite{Caffau12} is shown in \textsl{red}. The $\ion{Ti}{II}$ at $3361.2\,\AA$ was excluded from the $\chi^2$-analysis.} 
\label{fig:NH}
  \resizebox{3.5in}{!}{\includegraphics{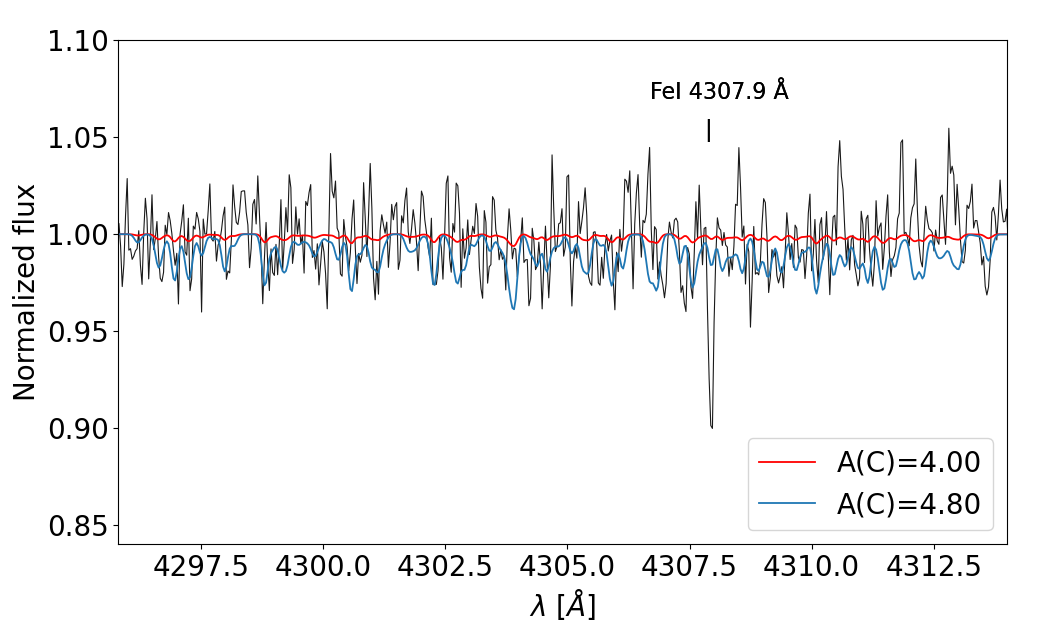}}
\caption[]{CH G-band at $4300\,\AA$. The $\ion{Fe}{I}$ at $4307.9\,\AA$ was excluded from the $\chi^2$-analysis.} 
\label{fig:CH}
\end{figure}

\begin{figure}[ht]
  \centering
  \resizebox{3.5in}{!}{\includegraphics{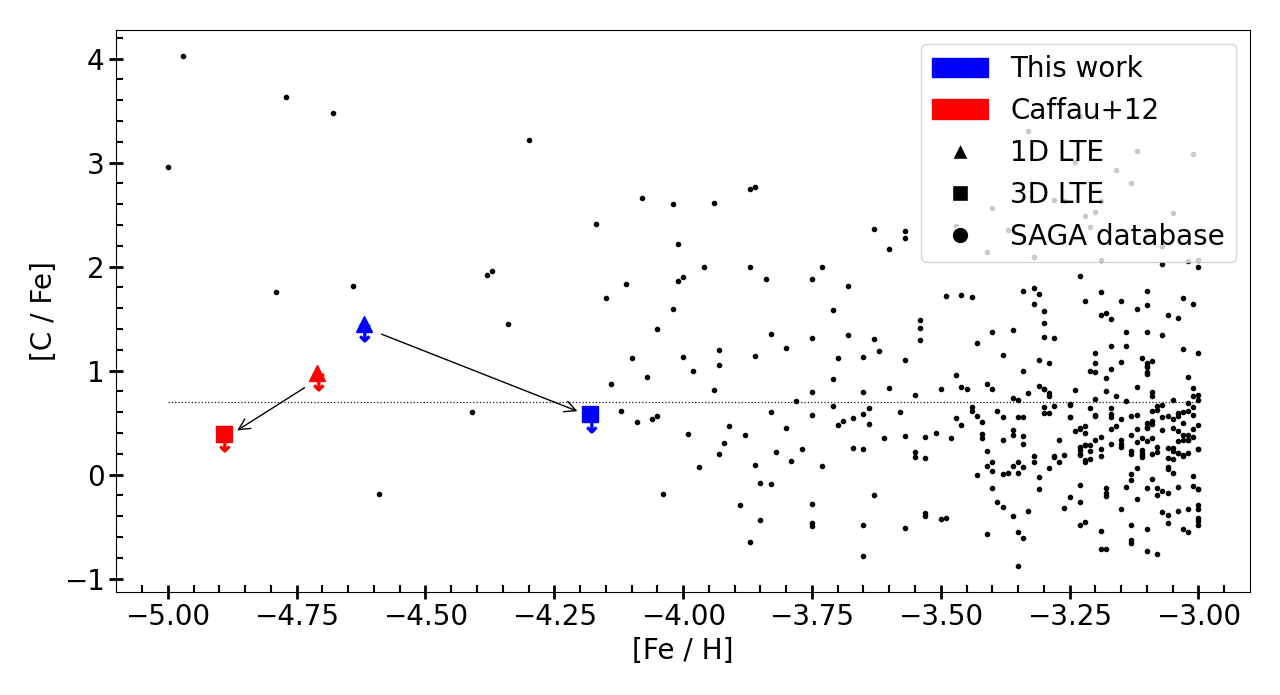}}
\caption[]{Upper limits on $[\mathrm{C}/\mathrm{Fe}]$ for SDSS J102915+172927 in 1D LTE (\textsl{triangles}) and 3D LTE (\textsl{squares}) for both the original limits \citep[red]{Caffau12}, and the results of this work (\textsl{blue}). For comparison, we show the $[\mathrm{C}/\mathrm{Fe}]$ abundance for a sample of metal-poor stars from the SAGA database \citep[\textsl{black circles}]{Suda08} and the cutoff for carbon-enhancement at $[\mathrm{C}/\mathrm{Fe}]=0.7$ (\textsl{dotted line}) from \cite{Aoki07}} 
\label{fig:CoverFe}
\end{figure}

Using this method, we increase the upper limit significantly for both molecular bands: $A(\mathrm{C}) = 4.86\,\mathrm{dex}$ and $A(\mathrm{N}) = 4.65\,\mathrm{dex}$, as compared to \cite{Caffau12}: $A(\mathrm{C})_\mathrm{Caffau} = 3.96\,\mathrm{dex}$ and $A(\mathrm{N})_\mathrm{Caffau} = 2.83\,\mathrm{dex}$. The corresponding synthetic spectra are shown in Figs.~\ref{fig:NH} and \ref{fig:CH} for N and C, respectively. In addition, the new and old upper limits on C are shown together with a subset of metal-poor stars from the SAGA database \citep{Suda08} in Fig.~\ref{fig:CoverFe}.

The large increase of the C and N upper limit can originate from multiple sources such as: changes in surface gravity, spectrum synthesis code, and atmosphere between the original and this work. In the case of the CH and NH upper limits, however, we believe that the main discrepancy is likely due to a difference in methodology when determining these limits. \citet{Caffau12} show in their Figs. 6 and 7 the molecular bands for different abundances $A(\mathrm{C})=4.65,\,6.00\,\mathrm{and}\,A(\mathrm{N})=3.60,\,5.40$. From visual inspection, their $A(\mathrm{C})=5.40$ spectrum corresponds well to our synthetic spectrum with $A(\mathrm{C})=5.00$, which points towards a $\sim0.4\,\mathrm{dex}$ abundance difference originating from the difference in surface gravity, atmospheric model and spectrum synthesis code. We hypothesise that the remaining difference is due to a difference in methodology. It is not described in \cite{Caffau12}, however, how the upper limits are derived such that a more direct comparison can not be made.

\subsection{Sodium and aluminium}
No Na or Al absorption features could be identified in the spectrum. We therefore present upper limits for the two elements, in this star for the first time, at the \ion{Na}{I} line at $5891.6\,\AA$ and \ion{Al}{I} line at $3961.5\,\AA$ using the same method as for Li. The resulting $3\sigma$-upper limits are shown in Table~\ref{t:abundances}.

\subsection{Alpha elements}
\subsection*{Magnesium}
We measured four \ion{Mg}{I} lines which combine to a mean 3D non-LTE abundance of $A(\mathrm{Mg}) = 3.18\pm0.05$ that is marginally consistent with \cite{Caffau12}, $A(\mathrm{Mg})=3.06 \pm 0.12$. \cite{Sitnova19} computed a 1D non-LTE value of $A(\mathrm{Mg})=3.17 \pm 0.07$ using the updated surface gravity of $\log g=4.7$. 

\subsection*{Silicon}
One \ion{Si}{I} line was detected at $3905.5\,\AA$ for which we computed a 3D non-LTE abundance of $A(\mathrm{Si})=3.42\pm0.12$ that is smaller as compared to \cite{Caffau12}, $A(\mathrm{Si})=3.55\pm0.1$.

\subsection*{Calcium}
One \ion{Ca}{I} line was detected at $4226\,{\AA}$ together with the \ion{Ca}{II} H$\&$K lines and the \ion{Ca}{II} triplet at $\approx 8500\,{\AA}$. The \ion{Ca}{II} H$\&$K and the weakest line of the \ion{Ca}{II} triplet were discarded due to being blended or too weak for a reliable equivalent width measurement. Using the resulting lines, we are able to recover ionisation balance between \ion{Ca}{I} and \ion{Ca}{II} in 3D non-LTE: $A(\mathrm{Ca})_\ion{Ca}{I}-A(\mathrm{Ca})_\ion{Ca}{II}=0.01\,\mathrm{dex}$ with a mean Ca abundance of $A(\mathrm{Ca}) = 1.87\pm0.02$. 

\cite{Caffau12} obtained a 3D + non-LTE abundance of $A(\mathrm{Ca})_\ion{Ca}{I} = 1.76\,\mathrm{and}\,A(\mathrm{Ca})_\ion{Ca}{II} = 1.35$. Using an updated model atom, \cite{Sitnova19} computed a 1D non-LTE Ca abundance, using similar stellar parameters as in this work, of $A(\mathrm{Ca})=1.82\pm0.06$, where they fixed the \ion{Ca}{I}/\ion{Ca}{II} ionisation equilibrium to derive the surface gravity.

\subsection*{Titanium}
We detected five \ion{Ti}{II} lines in the UVES spectrum of which three were deemed strong enough to determine a reliable equivalent width. The resulting mean 3D LTE abundance $A(\ion{Ti}{}) = 0.62\pm0.06$ is $+0.55\,\mathrm{dex}$ is higher than the original value derived by \cite{Caffau12}. A similar increase is seen in the 1D LTE case for which we obtain an abundance this is larger by $+0.39\,\mathrm{dex}$. Following Table~\ref{t:AbundCorrections} we can attribute approximately $0.2\,\mathrm{dex}$ to the change in surface gravity. The remaining $0.19\,\mathrm{dex}$ difference for the 1D LTE abundance can not be attributed to the different oscillator strengths used in this work, which are higher than the ones reported in \citet[Table 4]{Caffau12}, leaving the origin of the discrepancy unknown.

\subsection{Iron}
The UVES spectrum contains 52 visible \ion{Fe}{I} lines but no \ion{Fe}{II} lines. For the abundance computation a cutoff in reduced equivalent width of $\mathrm{log}[W_\lambda/\lambda_0]<-4.95$ was used to remove lines that do not fall on the linear part of the curve-of-growth \citep{Vickers00}\footnote{Chapter ``Stars, Spectroscopy of'', p.2199-2204, Figure 2}. The remaining 46 \ion{Fe}{I} lines are shown in Fig.~\ref{fig:Fe_excit} to the left of the solid black line. For the 1D LTE case we expect the only difference between our results and the one from \cite{Caffau12} to be the change in gravity ($\Delta\log g = +0.7\,\mathrm{dex}$), as well as differences between the spectral synthesis codes and the model atmosphere employed. We obtain a value of $A(\mathrm{Fe})=2.80\pm0.15$ that is comparable to the value found by \cite{Caffau12}, $A(\mathrm{Fe})=2.87\pm0.13$. This result is assuring since \ion{Fe}{I} is not strongly sensitive to surface gravity (see Table~\ref{t:AbundCorrections}).

Our 1D non-LTE abundance, $A(\mathrm{Fe})=3.05\pm0.13$, is comparable within its error to \cite{Caffau12}, $A(\mathrm{Fe})=3.00\pm0.13$. In addition, \cite{Ezzeddine17} recomputed the 1D non-LTE Fe abundance for this star, using a surface gravity of $\log g=4.0$, and obtained a slightly higher value of $A(\mathrm{Fe})=3.23\pm0.14$. They based their analysis only on three \ion{Fe}{I} lines, however.

From our consistent 3D non-LTE analysis we derive an abundance of $A(\mathrm{Fe})=3.28\pm0.13$ that is $0.57\,\mathrm{dex}$ higher than the 3D LTE + non-LTE value obtained by \cite{Caffau12}, $A(\mathrm{Fe})=2.71\pm0.10$. We emphasise that \cite{Caffau12} computed separate 1D non-LTE and 3D LTE abundance corrections and added these together to get a final 3D+non-LTE abundance. Applying a similar method to our abundances would yield a value of $A(\mathrm{Fe})_\mathrm{3D+NLTE}=2.97\pm0.14$, which has a better agreement with \cite{Caffau12}, but significantly lower than the fully consistent 3D non-LTE result. Hence, this work demonstrates again that 3D and non-LTE effects influence each other non-linearly and cannot be simply added together for metal-poor stars.  

This result supports the claim by \cite{Amarsi16b} that metallicities of the most-metal poor stars are often systematically underestimated. This is because for \ion{Fe}{I} lines, the non-LTE effects and the 3D effects tend to go in the same direction, and can enhance each other in warm metal-poor stars.

\subsection{Nickel}
We detected nine usable \ion{Ni}{I} lines in the UVES spectrum for which we computed a mean 3D LTE abundance of $A(\mathrm{Ni}) = 1.67\pm0.14$. This value is $+0.34\,\mathrm{dex}$ higher than the original value. On the other hand, our 1D LTE abundance agrees to within $+0.03\,\mathrm{dex}$ with the value of \cite{Caffau12}, $A(\mathrm{Ni}) = 1.65\pm0.14$.

\begin{table*}[t]
\begin{center}
    \caption{Summary of the computed abundances ($A(\mathrm{X})$, see Eq.~\ref{eq:Ax}) and upper limits together with 3D+non-LTE values from \cite{Caffau12}. } 
    \def\arraystretch{1.5}
    \begin{tabular}{ cccccc|c}
     \hline
     \hline
     Species & 1D LTE & 1D non-LTE & 3D LTE & 3D non-LTE & $[\mathrm{X}/\mathrm{Fe}]_\mathrm{3D\,non-LTE}$ & $\mathrm{Caffau}_\mathrm{3D + non-LTE}$\\ \hline
      $\ion{Mg}{I}$ & $2.90 \pm 0.10$ & $3.09 \pm 0.09$ & $2.93 \pm 0.11$ & $3.18 \pm 0.11$ & $-0.19$ & $3.05\pm0.12$  \\
      $\mathrm{\ion{Ca}{I}}$ & $1.58 \pm 0.13$ & $1.72 \pm 0.12$ & $1.57 \pm 0.10$ & $1.88 \pm 0.08$ & $-0.24$ & $1.79\pm0.10$\\
      $\mathrm{\ion{Ca}{II}}$ & $2.14 \pm 0.13$ & $1.87 \pm 0.14$ & $1.96 \pm 0.12$ & $1.87 \pm 0.11$ & $-0.25$ & $1.38\pm0.09$\\
      $\ion{Si}{I}$ & $3.16 \pm 0.13$ & $3.24 \pm 0.12$ & $3.20 \pm 0.12$ & $3.42 \pm 0.12$ & $0.09$ & $3.56\pm0.10$\\
      $\ion{Fe}{I}$ & $2.80 \pm 0.22$ & $3.05 \pm 0.20$ & $2.70 \pm 0.20$ & $3.28 \pm 0.19$ & $\dots$ & $2.63\pm0.10$ \\
      $\ion{Ti}{II}$ & $0.54 \pm 0.18$ & $\dots$ & $0.62 \pm 0.13$ & $\dots$ & $ -0.17 $ & $0.07\pm0.16$\\
      $\ion{Ni}{I}$ & $1.71 \pm 0.22$ & $\dots$ & $1.67 \pm 0.20$ & $\dots$ & $-0.38$ & $1.33\pm0.11^1$\\ \hline
      $\mathrm{C}$ & $<5.25\pm0.14$ & $\dots$ & $<4.86\pm0.15 $ & $\dots$ & $<0.58$ & $<3.96^1$\\
      $\mathrm{N}$ & $<5.13\pm0.18$ & $\dots$ & $<4.65\pm0.20$ & $\dots$ & $<1.00$ & $<2.83^1$\\
      $\ion{Li}{I},\,3\sigma$ & $<1.02\pm0.05$ & $<0.93\pm0.05$ & $<1.00\pm0.05$ & $<1.06\pm0.05$ & $<1.99$ & $<0.90$\\
      $\ion{Na}{I},\,3\sigma$ & $<1.55\pm0.10$ & $<1.48\pm0.10$ & $<1.58\pm0.15$ & $<1.62\pm0.15$ & $<-0.42$ & $\dots$ \\
      $\ion{Al}{I},\,3\sigma$ & $<1.51\pm0.10$ & $<1.90\pm0.10$ & $<1.51\pm0.15$ & $<2.05\pm0.15$ & $<-0.20$ & $\dots$ \\
     \hline

    \end{tabular}
\label{t:abundances}
\end{center}
\footnotesize{$^1$ 3D LTE abundance from \cite{Caffau12}}\\
\end{table*}

\section{Discussion}\label{Sec:discussion}
With our newly derived 3D non-LTE abundances, we can now address some of the questions regarding the properties of Pop III stars and low-mass star formation introduced in $\S$\ref{Sec:introduction}. 

\subsection{Supernova yield comparison}
We compared our 3D non-LTE abundances with yields of core-collapse supernovae of non-rotating Pop III stars \cite{Heger10}, updated in 2012 for missing low-mass models\footnote{Data is available at \url{https://starfit.org}}, using the \texttt{StarFit} code \citep{Heger23}. Specifically, we used the Python version (\texttt{0.17.12}) of \texttt{StarFit}\footnote{Data and routines are available at \url{https://pypi.org/project/starfit/}} to find the best match between an input abundance pattern and supernova model employing a reduced $\chi^2$-algorithm ($\Bar{\chi}^2$) \citep[see their Section 7]{Heger10}.  Upper limits, however, are now treated by \texttt{StarFit} as in Eq.~(6) of \citet{Magg20} --- they contribute to $\bar\chi^2$ similar to a measurement mismatch if violated and are basically ignored when well fulfilled.. The one-dimensional mixing $\&$ fallback models (`\texttt{*.S4.*}') cover the following range of progenitor star masses $9.6$--$100\,\mathrm{M}_\sun$, as well as piston-driven explosion energies $E_{51} = 0.3$--$10$ (with $E_{51} = 10^{51}$\,erg) and hydrodynamic mixing fractions $f_\mathrm{mix} = 0$--$0.25$ (see \citealt{Heger10} for details).  In addition, the fitting code employs a dilution factor $f_\mathrm{dil}$ that accounts for mixing of the supernova ejecta with the interstellar medium (Big Bang composition). This free parameter essentially shifts the whole supernova abundance pattern up or down. Li is excluded from the fitting procedure and C, N, Al, and Na are treated as upper limits. For Na and Al, more strict $1\sigma$ upper limits are calculated with Cayrel's formula \citep{Cayrel88,Cayrel04}, to use in the supernova fitting: $A(\mathrm{Na})_{1\sigma}<1.12\pm0.15$ and $A(\mathrm{Na})_{1\sigma}<1.54\pm0.15$.

\begin{figure}[t]
  \centering
  \resizebox{3.2in}{!}{\includegraphics{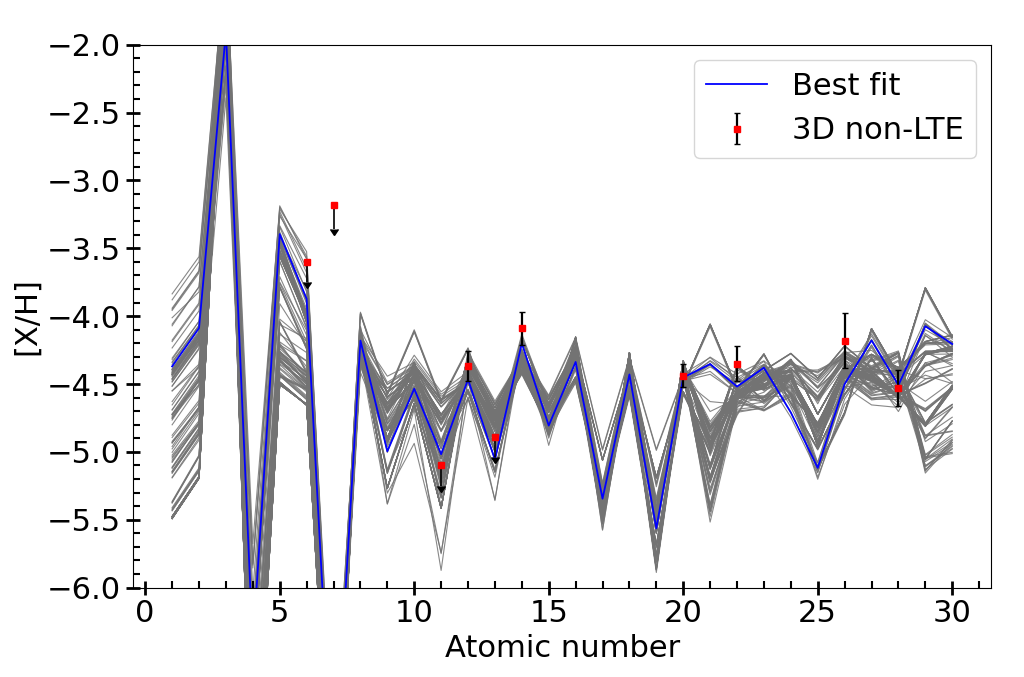}}
  \resizebox{3.2in}{!}{\includegraphics{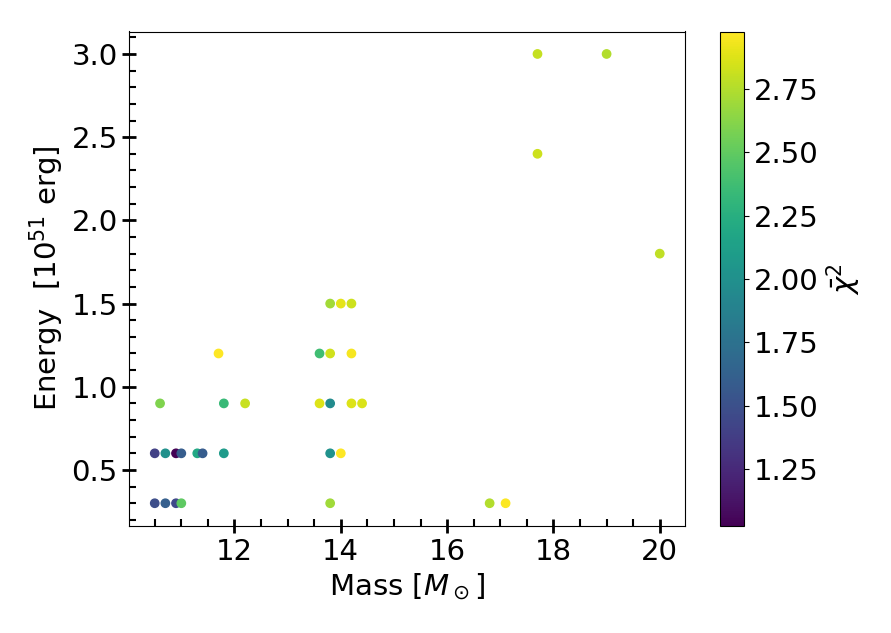}}
  \caption[]{The top panel shows the 3D non-LTE abundances (\textsl{red squares}) together with all supernovae models that have $\Bar{\chi}^2<3$ (\textsl{solid grey lines}), the best fit is shown in \textsl{blue}. The \textsl{Bottom Panel} show the progenitor mass and explosion energy for the same best-fitting models as shown in the \textsl{Top Panel}, together with their respective $\Bar{\chi}^2$ values.} 
\label{fig:massVSenergy}
\end{figure}

The best fitting model has a progenitor mass of $M = 10.9\,\mathrm{M}_\sun$, an explosion energy of $E_{51}=0.6 $ and a mixing efficiency of $f_\mathrm{mix}=0.01$, and matches with $\Bar{\chi}^2_\mathrm{best}=1.03$. The variance of the best-fitting models ($\Bar{\chi}^2<3$) was analysed by selecting for each mass-energy pair the $f_\mathrm{mix}$ that has the lowest $\Bar{\chi}^2$. The reasoning is that mixing acts as a free parameter in the one-dimensional explosion models. Moreover, the effect of mixing is small at masses $\sim10\,\mathrm{M}_\sun$ and mainly affects elements with $Z>20\,(\mathrm{Ca})$. The resulting distribution in mass and energy, and corresponding abundance patterns are shown in Fig.~\ref{fig:massVSenergy}. We find that only a narrow range of progenitor masses, $M=10-20\,\mathrm{M}_\sun$, and explosion energies, $E_{51} = 0.3-3$, are able to fit the abundance pattern of SDSS J102915+172927.

\begin{figure}[t]
  \centering
  \resizebox{3.5in}{!}{\includegraphics{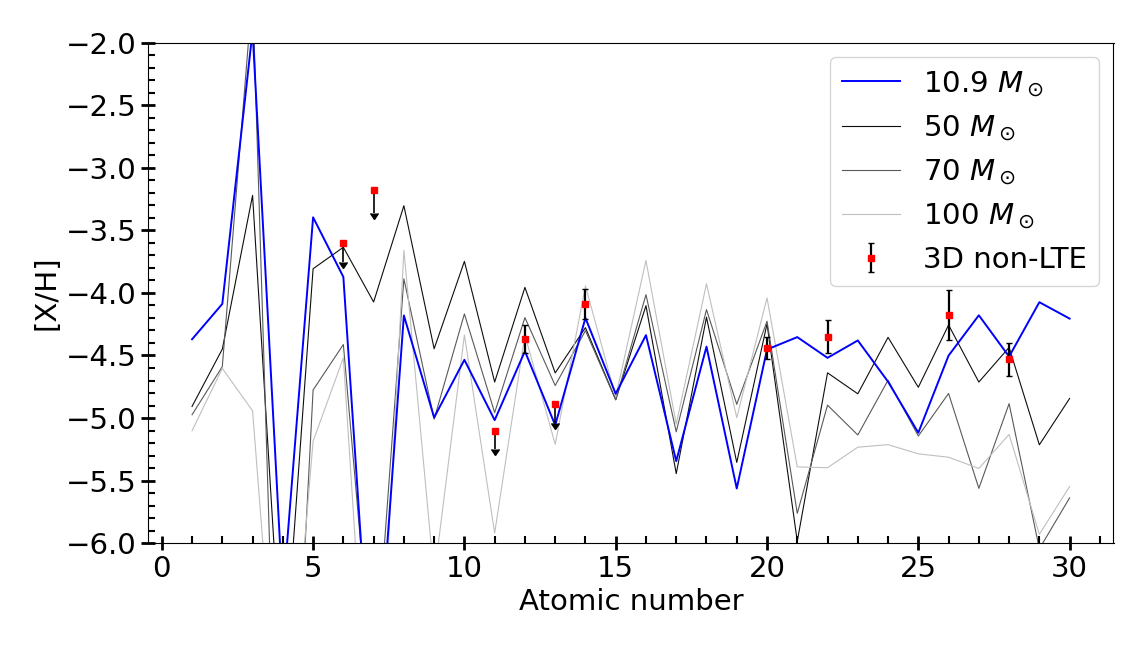}}
\caption[]{3D non-LTE abundances from this paper (\textsl{red squares}) are shown together with the best fitting progenitor supernova-model in \textsl{blue} and three other models with a higher progenitor mass (\textsl{grey}): $M=50$, $70$, $100\,\mathrm{M}_\sun,\,E_{51}=0.3$ and $f_\mathrm{mix}=0.25$.} 
\label{fig:bestSNe}
\end{figure}

The difference between the best fit and yields from models with higher progenitor masses is shown in Fig. \ref{fig:bestSNe}. At higher masses the supernova underproduces heavy elements (Ti, Fe, Ni) and overproduces Mg, Si, and Ca compared to the best fit model.

Applying the same fitting routine to our 1D LTE abundances results in a broader mass $M\approx10-30$ and energy range $E_{51} = 0.3-10$ for all models with $\Bar{\chi}^2<3$. This is explained by the lower 1D LTE Fe abundance that allows for models with higher progenitor mass to achieve good fits. Already in Fig.~\ref{fig:bestSNe} it is noticeable that most supernova yields have lower Fe abundances than our 3D non-LTE result.

After the original discovery of SDSS J102915+172927, there have been several papers comparing its abundance pattern to theoretical yields of Pop III supernovae. The following literature results all use the original abundances from \cite{Caffau12}. Similar to this work, \cite{Placco15} used the supernova models of \cite{Heger10} to find a best fitting progenitor mass of $M=10.6\,\mathrm{M}_\sun$ with explosion energy $E_{51}=0.9$, which is in agreement with our work. \cite{Schneider12b} used the Pop III core-collapse supernova models by \cite{Limongi12} to find a best fit for progenitors with mass $M=20~\mathrm{and}~35~\mathrm{M}_\odot$ and explosion energy $E_{51}=1$. \cite{Tominaga14} compared three core-collapse SNe models taken from \cite{Iwamoto05,Tominaga07b} with varying mass ($25\,\mathrm{and}\,40\,\mathrm{M}_\sun$) and mixing efficiency with a sample of metal-poor stars including SDSS J102915+172927. Their best match is a progenitor star with $M = 25\,\mathrm{M}_\sun$ without mixing enhancement. Using the same Pop III supernova and hypernova models as \cite{Tominaga14}, \cite{Ishigaki14} found a better agreement with a hypernova of $M=40\,\mathrm{M}_\sun$, $E_{51}=30$ and mixing $\mathrm{log}f=-0.9$, where they attribute the obtained explosion energy to the relatively high [Si/Ca] ratio. These papers find a higher progenitor mass and explosion energy than this work. We note, however, that the data sets of \cite{Iwamoto05} and \cite{Tominaga07b} only cover a comparably limited progenitor mass range.
In summary, based on our new analysis and using the model set of \citet{Heger10}, we find that the observed abundance pattern of SDSS J102915+172927 can be well explained by a typical single star supernova (typical mass range $10$--$20\,\mathrm{M}_\odot$ and typical explosion energy, $E_{51} = 0.3-3$) of a compact Pop III star (low mixing, $f_\mathrm{mix}=0.01$, as compared to red super-giants, consistent with the hydrodynamical models of \citealt{Joggerst2009}). 

\subsection{Mean alpha abundance}
\begin{figure}[ht]
  \centering
  \resizebox{3.2in}{!}{\includegraphics{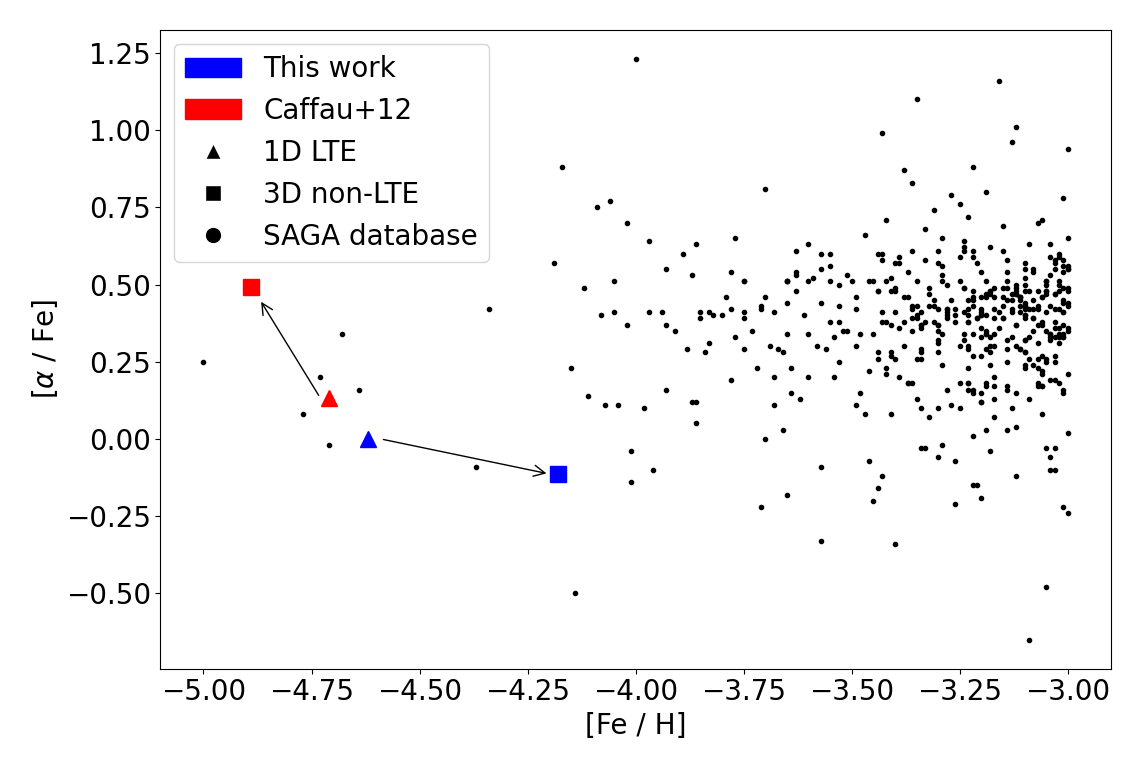}}
\caption[]{Similar to Fig.~\ref{fig:CoverFe} but now for the mean $[\alpha/\mathrm{Fe}]$ abundance.} 
\label{fig:alphaoverFe}
\end{figure}

Using our new 3D non-LTE abundances, we computed the mean alpha-abundance using a simple mean of: $[\mathrm{Si}/\mathrm{Fe}]$, $[\mathrm{Mg}/\mathrm{Fe}]$, and $[\mathrm{Ca}/\mathrm{Fe}]$. We find that the mean alpha enhancement of SDSS J102915+172927 decreases substantially compared to the original work:
\begin{equation}
    [\alpha/\mathrm{Fe}]_\mathrm{Caffau}=0.49 \,\rightarrow\,[\alpha/\mathrm{Fe}]_\mathrm{Lagae} = -0.11\;,
\end{equation}
also shown in Fig.~\ref{fig:alphaoverFe}. The observed change from alpha-enhanced ($[\alpha/\mathrm{Fe}]\sim0.4$) to sub-solar alpha abundance is mainly driven by the relative large increase in Fe abundance, compared to the alpha-elements, when going from 1D LTE to 3D non-LTE. \citet{Amarsi22} found that for warm metal-poor subgiants ($[\mathrm{M}/\mathrm{H}]=-3$) the change in iron abundance, $A(\ion{Fe}{I})_\mathrm{1DLTE}-A(\ion{Fe}{I})_\mathrm{3DnonLTE}$, can be as large as $0.4-0.5\,\mathrm{dex}$. 3D non-LTE abundance changes for the alpha elements are expected to be smaller. Hence, it is not unrealistic to expect similar shifts in alpha-enhancement for other ultra metal-poor stars, shifting the mean alpha-to-iron ratio of the stars shown in Fig.~\ref{fig:alphaoverFe} downwards. This would imply that the yields of Pop III supernovae would have produced a lower $[\alpha/\mathrm{Fe}]$ than previously thought. Verification of this claim, however, would require a consistent 3D non-LTE abundance analysis of extremely-metal poor stars ($[\mathrm{Fe}/\mathrm{H}]<-3$).

Our atmospheric models where calculated using alpha elements enhanced by $\mathrm{[\alpha/Fe]}=+0.4$, contrary to our findings that SDSS J102915+172927 has a sub-solar alpha abundance. At these low metallicities H is expected to be the main electron donor such that small changes in metal mass fraction are not expected to significantly impact the atmosphere. To test this, we compared two \texttt{MARCS} models ($T_\mathrm{eff}= 5750~\mathrm{K}$, $\log g=4.5$, $\mathrm{[Fe/H]}=-2.5$, $v_{micro}=1 \mathrm{km/s}$) that are alpha enhanced ($\mathrm{[\alpha/Fe]}=+0.4$) and alpha normal ($\mathrm{[\alpha/Fe]}=+0.0$), resulting in maximum temperature differences of $15~\mathrm{K}$ throughout the atmosphere.

\subsection{Impact on Pop II star formation theories}
In the original work, \cite{Caffau12} concluded that the low carbon upper limit of SDSS J102915+172927 excluded metal-line cooling as a possible mechanism of its star formation. In addition, the total metal mass fraction made SDSS J102915+172927 the most-metal poor star known to date. Using the new 3D non-LTE abundances derived in this work, we first make an estimate of the metal mass fraction $Z$:
\begin{equation}
    \frac{Z}{X} = \sum_\mathrm{i=element}\Big[10^{A(i)-12}\cdot A_\mathrm{w}(i)\Big]\;,
\end{equation}
where $A(i)$ is the abundance and $A_\mathrm{w}$ the atomic weight of element $i$. Subsequently, the normalisation $X+Y+Z=1$ allows to compute the total metal mass fraction $Z$:
\begin{equation}\label{eq:Z}
    Z = \frac{1 - Y}{1+ \Bigg(\sum_\mathrm{i=element}\Big[ 10^{A(i)-12}\cdot A_\mathrm{w}(i)\Big]\,\Bigg)^{-1}} \,\,,
\end{equation}
where $Y = 0.2477$ is the primordial He mass fraction \citep{Peimbert07}. Using equation \ref{eq:Z}, we computed two estimates of the mass fraction, $Z$: one where we restricted the sum over all elements to the species that \cite{Caffau12} reported (including upper limits), to make a consistent comparison, and one that used the abundances from the best fitting Population III supernova model.

In the first case we included oxygen as $[\mathrm{O}/\mathrm{Fe}]=+0.6$ and Sr with the upper limit as derived in \citep{Caffau12}. This results in a total metal content of $Z\lessapprox1.38\cdot10^{-6}$ or $Z/\mathrm{Z}_\sun\lessapprox9.8\cdot10^{-5}$, that is approximately a factor two higher than the original estimation $Z_\mathrm{Caffau+12}/\mathrm{Z}_\sun=5\cdot10^{-5}$. In the second case, applying the abundances of all species from Li to Zn of the best fitting supernova model, we obtain a value of $Z_\mathrm{SNe}\approx9.1\cdot10^{-7}$ or $Z_\mathrm{SNe}/\mathrm{Z}_\sun\approx6.6\cdot10^{-5}$. Even though this estimate includes more elements, the significantly lower N abundance of the supernova model results in an overall lower mass fraction compared to the first case. 

Turning now to the possibility of low mass star formation through the mechanism of metal-line cooling. It is theorised \citep{Bromm03} that this will only occur if there is a sufficient amount of C and O available in the star-forming cloud. Indeed, metal-line cooling is able to produce low-mass stars if the transition discriminant \citep{Bromm03,Frebel07}: 
\begin{equation}\label{eq:Dtrans}
    D=\mathrm{log\Big[10^{[\mathrm{C}/\mathrm{H}]} + 0.3\cdot10^{[\mathrm{O}/\mathrm{H}]}\Big]}\,\,,
\end{equation}
is greater than $D\ge D_\mathrm{crit}=-3.5\pm0.2$. Using our new upper limit on C and $[\mathrm{O}/\mathrm{Fe}]=+0.6$, we obtain a transition discriminant $D\le-3.6\pm0.15$ that is marginally comparable to the critical value needed for metal-line cooling, contrary to the original result of \cite{Caffau11} $D\le-4.2$. Hence, our result cannot exclude either metal-line cooling or dust-induced fragmentation as the underlying mechanism in the formation of SDSS J102915+172927.

\section{Conclusions}\label{Sec:conlusion}
In this work we have performed a fully consistent 3D non-LTE abundance analysis of SDSS J102915+172927, known as the most-metal poor star. For this purpose we employed a tailored 3D atmospheric model using improved stellar parameters and up-to-date atomic data to calculate synthetic spectra. The primary outcome is that the resulting Fe-abundance is $+0.57\,\mathrm{dex}$ higher than the original value, showcasing the importance of performing consistent 3D non-LTE calculations for ultra metal-poor stars. 

The increase in metallicity, together with new upper limits on C and N, has important implications regarding progenitor Pop III properties and low-mass star formation in the early universe. First, we find that the upper limit of the total metal content of the star increases by a factor 2. Together with the increased carbon upper limit, we find that the abundances of the star is consistent with formation both through metal-line cooling and dust-induced fragmentation. Observations with higher $S/N$ are necessary to improve the upper limit on C and to draw stronger conclusions. Secondly, our new 3D non-LTE abundances provide stronger constraints on the mass of the progenitor Pop III star as well as its explosion energy, assuming that SDSS J102915+172927 is mono-enriched, as compared to a 1D LTE analysis. In particular, we find that progenitors with mass $M=10-20\,\mathrm{M}_\sun$ and explosion energy $E_{51}=0.3-3$ are able to reproduce the abundance pattern. The best fit is a Pop III progenitor with mass $M=10.9\,\mathrm{M}_\sun$ that exploded with energy $E_{51}=0.6$.  These are typical masses and explosion energies for core collapse supernovae \citep[e.g.,][]{Mueller2020} and no hypernova model is required. Lastly, the strong increase in $A(\mathrm{Fe})$ coupled with relatively smaller changes in alpha-element abundances changes the status of the star from alpha-enhanced to a star with sub-solar alpha abundance. 

Following the conclusion of \cite{Nordlander17a}, and in light of upcoming large scale surveys which will provide dozens of new ultra metal-poor stars, it is critical for future studies to apply full 3D non-LTE abundance calculations whenever possible.


\begin{acknowledgements}
We thank the anonymous referee for their comments, which have improved the manuscript. We gratefully thank Federico Sestito for providing updated stellar parameters for SDSS J102915.14+172927.9 using Gaia DR3 data. This work has made use of the VALD database, operated at Uppsala University, the Institute of Astronomy RAS in Moscow, and the University of Vienna. CL and KL acknowledge funds from the European Research Council (ERC) under the European Union’s Horizon 2020 research and innovation programme (Grant agreement No. 852977). KL also acknowledges funds from the Knut and Alice Wallenberg foundation. AMA gratefully acknowledges support from the Swedish Research Council (VR 2020-03940). LFRD acknowledges support from the Carlsberg Foundation (grant agreement CF19-0649). TTH acknowledges support from the Swedish Research Council (VR 2021-05556). Parts of this research were supported by the Australian Research Council (ARC) Centre of Excellence (CoE) for All Sky Astrophysics in 3 Dimensions (ASTRO 3D), through project number CE170100013.  AH wad supported, in part, by the ARC CoE for Gravitational Wave Discovery (OzGrave) project number CE170100004 and acknowledges software development support from Astronomy Australia Limited's ADACS scheme (Project IDs AHeger\_2022B, AHeger\_2023A).  This work is based on data obtained from the ESO Science Archive Facility. The computations were performed on resources provided by the Swedish National Infrastructure for Computing (SNIC) at UPPMAX under project SNIC2022/5-351 and PDC under project PDC-BUS-2022-4, and at the Centre for Scientific Computing, Aarhus: \url{http://phys.au.dk/forskning/cscaa/}.
\end{acknowledgements}

%
%

\bibliographystyle{aa} 
\bibliography{main.bib} 

\begin{appendix} 
\section{$\Bar{\chi}^2$-values of the upper limit determination}

\begin{figure}[ht]
  \centering
  \resizebox{3.2in}{!}{\includegraphics{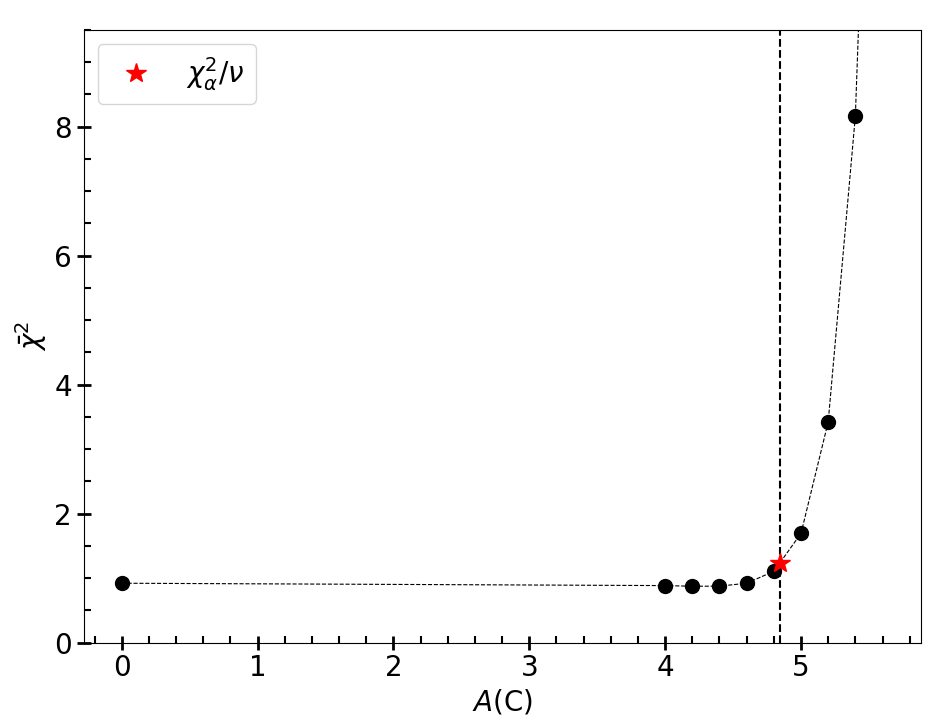}}
  \resizebox{3.2in}{!}{\includegraphics{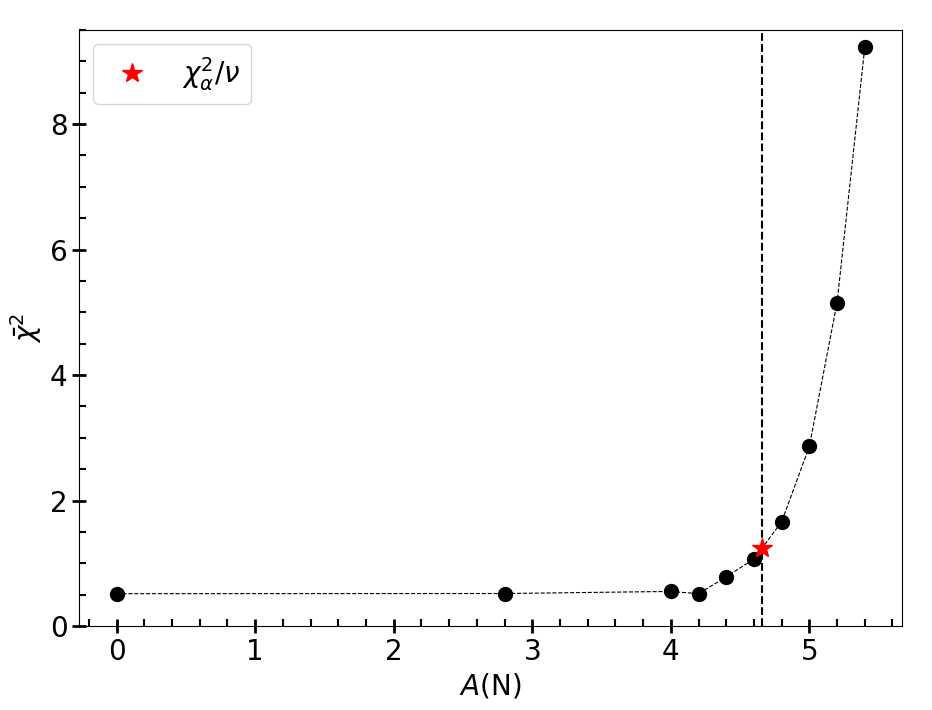}}
\caption[]{Overview of the $\Bar{\chi}^2$ values corresponding to the different synthetic spectra of CH and NH. The red star represents the interpolated abundance where $\Bar{\chi}^2\approx\chi^2_\alpha/\nu$, also highlighted with a vertical dashed line.} 
\label{fig:chi2_results}
\end{figure}

\section{Linelist}
\longtab[1]{
\begin{longtable}{lcrrr}
\caption{Equivalent widths of the measures spectral lines together with respective oscillator strength (log $gf$) and lower level excitation potential ($E_\mathrm{low}$). References to the atomic data are given in $\S$ \ref{Sec:spectralModelling}.}
\label{t:EWobs}
\def\arraystretch{1.5}
\\
\hline
\hline
     Ion & $\lambda$ & $W_\mathrm{UVES}$ & log $gf$ & $E_\mathrm{low}$\\
      & $[\AA]$ & $[$m\AA$]$ &  & $[\mathrm{eV}]$\\
     \hline
     \endfirsthead
     \caption{continued}\\
     \hline
     Ion & $\lambda$ & $W_\mathrm{UVES}$ & log $gf$ & $E_\mathrm{low}$\\
      & $[\AA]$ & $[$m\AA$]$ &  & $[\mathrm{eV}]$\\
     \hline
     \endhead
     \hline
     \endfoot
     \hline
     \endlastfoot
     $\ion{Li}{I}$ & 6707.6000 & $22\pm2$ & $0.174$ & 0.000 \\
     $\ion{Na}{I}$ & 5891.6000 & $22\pm2$ & $-0.194$ & 0.000   \\
     $\ion{Mg}{I}$ & 3832.29\phantom{00} & $28\pm4$ & $-0.339$ & 2.712  \\
                   & ... & ... & $0.138$ & ... \\
     $\ion{Mg}{I}$ & 3838.29\phantom{00} & $36\pm4$ & $-1.515$ & 2.717  \\   
     ... & ... & ... & $0.409$ & ...  \\   
     ... & ... & ... & $-0.339$ & ...  \\   
     $\ion{Mg}{I}$ & 5172.7140 & $11\pm1$ & $-0.363$ & 2.712  \\
     $\ion{Mg}{I}$ & 5183.6340 & $18\pm2$ & $-0.168$ & 2.717  \\
     $\ion{Al}{I}$ & 3961.5000 & $22\pm2$ & $-0.333$ & 0.014  \\
     $\ion{Si}{I}$ & 3905.5460 & $17\pm2$ & $-1.040$ & 1.909  \\
     $\ion{Ca}{I}$ & 4226.7500 & $22\pm2$ & $0.244$ & 0.000  \\
     $\ion{Ca}{II}$& 8542.0900 & $93\pm3$ & $-0.463$ & 7.813  \\
     $\ion{Ca}{II}$& 8662.1410 & $77\pm3$ & $-0.723$ & 7.806  \\
     $\ion{Ti}{II}$ & 3372.7926 & $36\pm9$  & 0.280 & 0.012  \\
     $\ion{Ti}{II}$ & 3361.2121 & $37\pm8$  & 0.410 & 0.028  \\
     $\ion{Ti}{II}$ & 3349.4022 & $39\pm11$ & 0.540 & 0.049   \\
     $\ion{Fe}{I}$ & 4383.5444 & $31\pm2$ & $0.208$  & 0.859  \\
     $\ion{Fe}{I}$ & 4325.7615 & $11\pm2$ & $0.006$  & 1.608 \\
     $\ion{Fe}{I}$ & 4307.9016 & $12\pm2$ & $-0.072$ & 1.557  \\
     $\ion{Fe}{I}$ & 4271.7599 & $18\pm2$ & $-0.173$ & 1.485  \\
     $\ion{Fe}{I}$ & 4202.0288 & $10\pm2$ & $-0.689$ & 1.485 \\
     $\ion{Fe}{I}$ & 4071.7375 & $13\pm2$ & $-0.008$ & 1.608  \\
     $\ion{Fe}{I}$ & 4063.5936 & $20\pm2$ & $0.062$  & 1.557  \\
     $\ion{Fe}{I}$ & 4045.8119 & $28\pm2$ & $0.284$  & 1.485 \\
     $\ion{Fe}{I}$ & 3930.2963 & $26\pm2$ & $-1.491$ & 0.087  \\
     $\ion{Fe}{I}$ & 3927.9194 & $19\pm2$ & $-1.522$ & 0.110  \\
     $\ion{Fe}{I}$ & 3922.9112 & $19\pm2$ & $-1.626$ & 0.052  \\
     $\ion{Fe}{I}$ & 3920.2574 & $8\pm2$  & $-1.734$ & 0.121  \\
     $\ion{Fe}{I}$ & 3899.7070 & $19\pm2$ & $-1.515$ & 0.087  \\
     $\ion{Fe}{I}$ & 3895.6559 & $14\pm2$ & $-1.668$ & 0.110  \\
     $\ion{Fe}{I}$ & 3886.2818 & $39\pm2$ & $-1.055$ & 0.052 \\
     $\ion{Fe}{I}$ & 3878.5728 & $32\pm2$ & $-1.379$ & 0.087  \\
     $\ion{Fe}{I}$ & 3878.0177 & $10\pm2$ & $-0.896$ & 0.958 \\
     $\ion{Fe}{I}$ & 3859.9110 & $61\pm2$ & $-0.698$ & 0.000  \\
     $\ion{Fe}{I}$ & 3856.3711 & $26\pm2$ & $-1.280$ & 0.052  \\
     $\ion{Fe}{I}$ & 3841.0475 & $11\pm2$ & $-0.044$ & 1.608  \\
     $\ion{Fe}{I}$ & 3840.4372 & $17\pm2$ & $-0.497$ & 0.990  \\
     $\ion{Fe}{I}$ & 3834.2221 & $15\pm2$ & $-0.269$ & 0.958 \\
     $\ion{Fe}{I}$ & 3827.8222 & $12\pm2$ & $0.094$ & 1.557  \\
     $\ion{Fe}{I}$ & 3825.8805 & $36\pm2$ & $-0.024$ & 0.915  \\
     $\ion{Fe}{I}$ & 3824.4432 & $29\pm2$ & $-1.342$ & 0.000  \\
     $\ion{Fe}{I}$ & 3820.4249 & $49\pm2$ & $0.157$ & 0.859  \\
     $\ion{Fe}{I}$ & 3815.8396 & $23\pm2$ & $0.237$ & 1.485  \\
     $\ion{Fe}{I}$ & 3812.9642 & $5.3\pm1.5$ & $-1.047$ & 0.958  \\
     $\ion{Fe}{I}$ & 3787.8799 & $11\pm2$ & $-0.838$ & 1.011  \\
     $\ion{Fe}{I}$ & 3767.1914 & $12\pm2$ & $-0.382$ & 1.011  \\
     $\ion{Fe}{I}$ & 3763.7886 & $27\pm2$ & $-0.221$ & 0.990  \\
     $\ion{Fe}{I}$ & 3758.2326 & $37\pm4$ & $-0.005$ & 0.958  \\
     $\ion{Fe}{I}$ & 3749.4848 & $32\pm4$ & $0.190$  & 0.915  \\
     $\ion{Fe}{I}$ & 3748.2618 & $30\pm4$ & $-1.008$ & 0.110 \\
     $\ion{Fe}{I}$ & 3745.8991 & $21\pm4$ & $-1.336$ & 0.121  \\
     $\ion{Fe}{I}$ & 3745.5608 & $50\pm4$ & $-0.767$ & 0.087  \\
     $\ion{Fe}{I}$ & 3737.1312 & $57\pm4$ & $-0.572$ & 0.052  \\
     $\ion{Fe}{I}$ & 3727.6187 & $17\pm4$ & $-0.601$ & 0.958  \\
     $\ion{Fe}{I}$ & 3722.5627 & $26\pm4$ & $-1.280$ & 0.087 \\
     $\ion{Fe}{I}$ & 3719.9344 & $63\pm4$ & $-0.424$ & 0.000 \\
     $\ion{Fe}{I}$ & 3705.5654 & $23\pm4$ & $-1.321$ & 0.052   \\
     $\ion{Fe}{I}$ & 3647.8422 & $32\pm3$ & $-0.141$ & 0.915  \\
     $\ion{Fe}{I}$ & 3631.4629 & $33\pm3$ & $0.0001$ & 0.958  \\
     $\ion{Fe}{I}$ & 3618.7675 & $35\pm4$ & $-0.003$ & 0.990  \\
     $\ion{Fe}{I}$ & 3608.8589 & $28\pm4$ & $-0.090$ & 1.011  \\
     $\ion{Fe}{I}$ & 3586.9843 & $14\pm3$ & $-0.795$ & 0.990  \\
     $\ion{Fe}{I}$ & 3581.1927 & $62\pm4$ & $0.415$  & 0.859  \\
     $\ion{Fe}{I}$ & 3565.3787 & $26\pm4$ & $-0.133$ & 0.958  \\
     $\ion{Fe}{I}$ & 3490.5735 & $30\pm4$ & $-1.105$ & 0.052  \\
     $\ion{Fe}{I}$ & 3475.4499 & $29\pm4$ & $-1.054$ & 0.087  \\
     $\ion{Fe}{I}$ & 3465.8603 & $26\pm5$ & $-1.192$ & 0.110   \\
     $\ion{Fe}{I}$ & 3440.9883 & $53\pm5$ & $-0.958$ & 0.052   \\
     $\ion{Fe}{I}$ & 3440.6054 & $53\pm6$ & $-0.673$ & 0.000  \\
     $\ion{Ni}{I}$ & 3619.3910 & $20\pm4$ & $-0.137$  & 0.423   \\
     $\ion{Ni}{I}$ & 3524.5360 & $50\pm6$ & $-0.157$  & 0.025   \\
     $\ion{Ni}{I}$ & 3510.3320 & $11\pm4$ & $-0.807$  & 0.212   \\
     $\ion{Ni}{I}$ & 3492.9570 & $27\pm6$ & $-0.407$  & 0.109   \\
     $\ion{Ni}{I}$ & 3461.6540 & $37\pm9$ & $-0.517$  & 0.025   \\
     $\ion{Ni}{I}$ & 3458.4600 & $23\pm6$ & $-0.377$  & 0.212   \\
     $\ion{Ni}{I}$ & 3446.2590 & $28\pm7$ & $-0.547$  & 0.109   \\
     $\ion{Ni}{I}$ & 3414.7650 & $46\pm8$ & $-0.167$  & 0.025   \\
     $\ion{Ni}{I}$ & 3392.9860 & $31\pm8$ & $-0.677$  & 0.025   \\
     \hline
     CH & G-band 4300 &  &  &   \\
     NH & NH-band 3360 &  &  &   \\
     \hline
\end{longtable}
}
\end{appendix}

\end{document}